\documentclass[showpacs]{revtex4}
\usepackage{amsfonts}
\usepackage{amsmath}
\usepackage{amssymb}
\usepackage{graphicx}

\begin{document}

\title{Optical Stern-Gerlach effect via a single traveling-wave light}
\author{Haihu Cui$^{1}$ and Wenxi Lai$^{2}$}\email{wxlai@pku.edu.cn}
\affiliation{$^{1}$ Department of Building Engineering, Inner Mongolia Vocational and Technical College of Communications, Chifeng 024005, China}
\affiliation{$^{2}$ School of Applied Science, Beijing Information Science and Technology University, Beijing 100192, China}

\begin{abstract}
In this paper, we propose a simplified model of optical Stern-Gerlach effect based on coherent coupling between clock transition of alkaline-earth single atoms and a traveling-wave light. It is demonstrated that spin-orbit coupling induced chiral motion in atom deflection appears under the strong atom-light interaction. The strong optical driving removes perturbation from the Doppler effect and back action effect to access the coherent system. In this process, superposition of distant matter waves connected to the arbitrary distribution of atom internal state could be predicted, which is important for the realization of atom interferometry and quantum state operation. The influence from atom relaxation and atom-atom interactions are discussed. Basic conditions of experimental design are given in the end of this work.
\end{abstract}


\maketitle

\begin{center}
\textbf{1. Introduction}
\end{center}
In Stern-Gerlach experiment, the deflection of atomic beams in inhomogeneous magnetic fields reflects directional quantization due to the magnetic moments of atoms~\cite{Haken}. The Stern-Gerlach effect could play important rule in qubit generation~\cite{Nielsen} and quantum state estimation~\cite{Cifuentes}.

Optical Stern-Gerlach effect generally indicates atom orbital quantization in optical fields. It was pointed out early that optical field gradient can split atomic beams into different directions~\cite{Kazantsev} and the corresponding effects were observed experimentally~\cite{Moskowitz,Gould,Sleator}. These optical field gradients are mainly constructed with standing waves. It is theoretically demonstrated that, in the frame of standing wave quantization, two-photon interactions~\cite{Cook} and atom-light coupling without rotating wave approximation~\cite{Lembessis} lead to optical Stern-Gerlach effect. Such effect can be used to implement quantum nondemolition measurement~\cite{Tan}, quantum erasure~\cite{Chianello}, Quantum-trajectory analysis~\cite{Chough} and linear quantum computation~\cite{Popescu}.

Recently, optical Stern-Gerlach effect based on optical helicity gradients~\cite{Kravets} has been proposed. The optical helicity gradient plays the role of magnetic field gradient, which makes the liquid crystal occur displacements~\cite{Kravets}. Directions of displacements depend on right-handed or left-handed helicity of light in the liquid crystal. Existence of all-optical Stern-Gerlach effect has been predicted theoretically~\cite{Karnieli} and proved in experiment quite recently~\cite{Yesharim}. In the all-optical Stern-Gerlach effect, an idler beam is incident on a pumped nonlinear crystal, and is deflected into two mutual beams with discrete directions due to a transverse gradient in the nonlinear coupling.

In this paper, we investigate that a single traveling-wave light coupled to optical clock transition of neutral atoms could naturally induces optical Stern-Gerlach effect without the requirement of field gradient or standing waves. Although similar description has been given much earlier considering spontaneous emission of photons~\cite{Cook}, here we centered at coherent coupling between atom and light with both detuning and resonant interactions. We will analyze energy structure of atom internal states under optical driving and show how to remove noise originated from Doppler effect and back action effect. Experiments on coherent atom-light interactions can lead to spin-orbit coupling~\cite{Lin,Jimenez-Garcia} and topological chiral edge state~\cite{Curiel,Kanungo}, which attract intensive attentions recently. Ultra-long lifetime optical transitions could be provided from clock transitions of alkaline-earth (like) atoms. Lifetime of the alkaline-earth (like) atoms can be ranged from microsecond to tens of second~\cite{Ye,Livi}. Furthermore, Single-atom physics is still a significant area of research in the current information processing~\cite{Thomas,Chang,Wen-Xi}. Our present simplified model of optical Stern-Gerlach effect may have potential applications in quantum state detection~\cite{Cifuentes}, atom interferometry~\cite{Huesmann,Lachmann}, quantum information processing~\cite{Popescu,Byrnes,Lange}, where superposition of matter waves is important.

\begin{figure}
  \includegraphics[width=8 cm]{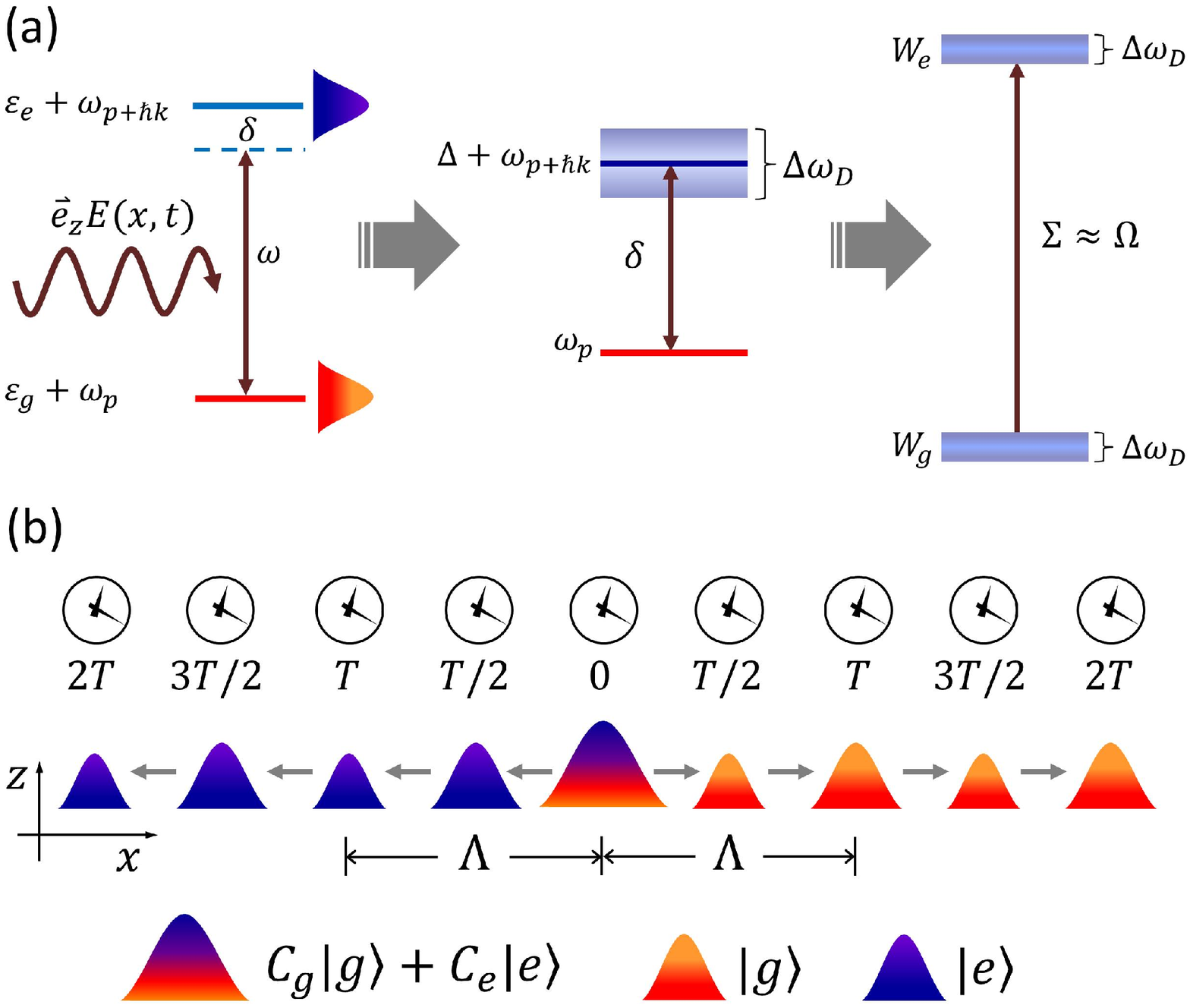}\\
  \caption{(Color on line) (a) From original energy structure to eigenfrequency structure. (b) Schematic illustration of wave packet splitting.}
  \label{fig1}
\end{figure}

\begin{center}
\textbf{2. Model}
\end{center}
Our model is a dilute atomic beam that consists of alkaline-earth metal atoms. Without loss of generality, we consider a Strontium ($^{87}$Sr) atom as a sample. This atom has long coherent time clock transition $^{1}$S$_{0}-^{3}$P$_{0}$~\cite{Campbell}. Considering the dilute gas in free-space, atom-atom repulsion would be neglected. The model that describing the coupling between an atom wave packet and a linearly polarized traveling light $\vec{\textbf{E}}(\textbf{x},t)=\vec{\textbf{e}}_{z}E_{0}cos(\omega t-k\textbf{x}+\phi_{0})$ is sketched in Fig.~\ref{fig1}. The corresponding Hamiltonian could be written as,
\begin{eqnarray}
&&\textbf{H}=\sum_{n=g,e}\varepsilon_{n}|n\rangle\langle n|+\int d p \frac{p^{2}}{2M}|p\rangle\langle p|-\frac{\hbar\Omega}{2}\int d p (e^{i\omega t}|g,p-\hbar k\rangle\langle e,p|+e^{-i\omega t}|e,p+\hbar k\rangle\langle g,p|),
\label{eq:total-Ham}
\end{eqnarray}
where $\varepsilon_{g}$ and $\varepsilon_{e}$ are two levels connecting the clock transition. $\Omega=|\mu|E_{0}/\hbar$ is the Rabi frequency~\cite{Meystre,Scully}. Initial phase $\phi_{0}$ of the light has been taken to be zero and the dipole moment $\mu$ is supposed to be real as they do not influence our following discussions.

Using the unitary transformation $e^{it\textbf{H}_{0}/\hbar}$ with $\textbf{H}_{0}=\varepsilon_{g}|g\rangle\langle g|+(\varepsilon_{g}+\hbar\omega)|e\rangle\langle e|$, we obtain a time independent Hamiltonian in interaction picture,
\begin{eqnarray}
&&\textbf{V}=\hbar\Delta|e\rangle\langle e|+\int d p \frac{p^{2}}{2M}|p\rangle\langle p|-\frac{\hbar\Omega}{2}\int d p (|g,p-\hbar k\rangle\langle e,p|+|e,p+\hbar k\rangle\langle g,p|),
\label{eq:Ham}
\end{eqnarray}
where $\Delta=\varepsilon_{e}-\varepsilon_{g}-\hbar\omega$ represents the detuning. Time evolution of the atom wave packet should be governed by the Schr\"{o}dinger equation in the interaction picture
\begin{eqnarray}
i\hbar \frac{\partial}{\partial t}|\varphi(t)\rangle=\textbf{V}|\varphi(t)\rangle.
\label{eq:S-equ}
\end{eqnarray}
Since, $|n,p\rangle$ is eigenvector of non-interacting Hamiltonian, the new wave function $|\varphi(t)\rangle$ in Eq.\eqref{eq:S-equ} can be expanded with the probability amplitude $\varphi_{n}(p,t)$ in momentum space as $|\varphi(t)\rangle=\int d p \sum_{n=g,e}\varphi_{n}(p,t)|n,p\rangle$.

\begin{figure}
  \includegraphics[width=8.5 cm]{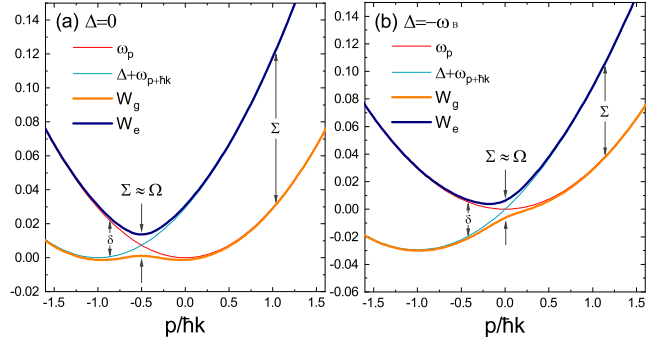}\\
  \caption{(Color on line) Scattering spectrum of the eigenfrequencies $W_{g}$, $W_{e}$ and the two levels $\omega_{p}$, $\Delta+\omega_{p+\hbar k}$ in interaction picture under the Rabi frequency $\Omega=2$ KHz. }
  \label{fig2}
\end{figure}

For a given momentum $p$, we have a sub-space $\{|g,p\rangle,|e,p+\hbar k\rangle\}$ in which Eq.\eqref{eq:S-equ} can be written in a $2\times2$ matrix form
\begin{eqnarray}
i\frac{\partial}{\partial t}\left[\begin{array}{c}
     \varphi_{g}(p,t) \\
     \varphi_{e}(p+\hbar k,t)
  \end{array}\right]=\left[\begin{array}{c c}
      \omega_{p} & -J \\
     -J & \Delta +\omega_{p+\hbar k}
  \end{array}\right]\left[\begin{array}{c}
     \varphi_{g}(p,t) \\
     \varphi_{e}(p+\hbar k,t)
  \end{array}\right],
\label{eq:S-equ2}
\end{eqnarray}
where $\omega_{p}=\frac{p^{2}}{2M\hbar}$ and $J=\frac{\Omega}{2}$. One can solve Eq.\eqref{eq:S-equ2} to obtain the general wave function with the superposition of dressed states,
\begin{eqnarray}
|\varphi(t)\rangle=\int dp (\varphi_{g}(p,t)|g,p\rangle+\varphi_{e}(p+\hbar k,t)|e,p+\hbar k\rangle).
\label{eq:solution1}
\end{eqnarray}

To solve Eq.\eqref{eq:S-equ2}, firstly the sub-space Hamiltonian could be diagonalized to achieve the eigenfrequencies, $W_{g,e}=\frac{1}{2}(\Delta+\omega_{p}+\omega_{p+\hbar k}\mp\Sigma)$. Here, we have the effective Rabi frequency $\Sigma=\sqrt{\delta^{2}+\Omega^{2}}$ with energy shift $\delta=\Delta+\omega_{D}+\omega_{B}$ which consists of the Doppler shift $\omega_{D}=\frac{p\hbar k}{M}$ and the back action (recoil energy) shift $\omega_{B}=\frac{\hbar k^{2}}{2M}$ due to atom-photon collision.

The area shown in Fig.~\ref{fig2} is called strong coupling regime where $\Omega\gg\delta$. In this regime, the energy shifts originated from the Doppler effect and the back action would be removed and one reach the result $\Sigma=W_{2}-W_{1}\approx \Omega$. In this condition, coherent periodic motion of atom in free-space could be concluded. Energy structures in Fig.~\ref{fig2} reveals chiral motion of particles related to spin-orbit couplings. Indeed, similar energy structures are also appeared in optical strong coupling system of Bose-Einstein condensates~\cite{Lin,Jimenez-Garcia,Curiel}, where spin-orbit couplings and topological chiral edge states are exploited.

Diagonalization of the sub-space Hamiltonian in Eq.\eqref{eq:S-equ2} also provides us eigenstates of this system. Then, the wave function amplitudes could be written with superposition of these eigenstates as
\begin{eqnarray}
\varphi_{g}(p,t)=(A_{+}e^{i\Sigma t/2}+A_{-}e^{-i\Sigma t/2})e^{-i(\Delta+\omega_{p+\hbar k}+\omega_{p})t/2},
\label{eq:amp1}
\end{eqnarray}
\begin{eqnarray}
\varphi_{e}(p+\hbar k,t)=(B_{+}e^{i\Sigma t/2}+B_{-}e^{-i\Sigma t/2})e^{-i(\Delta+\omega_{p+\hbar k}+\omega_{p})t/2},
\label{eq:amp2}
\end{eqnarray}
where, the coefficients are $A_{\pm}=\frac{1}{2\Sigma}[(\Sigma\pm\delta)\varphi_{g}(p,0)\pm\Omega\varphi_{e}(p+\hbar k,0)]$ and $B_{\pm}=\frac{1}{2\Sigma}[(\Sigma\mp\delta)\varphi_{e}(p+\hbar k,0)\pm\Omega\varphi_{g}(p,0)]$. The initial state amplitude of the atom is composed of two parts, $\varphi_{n}(p,0)=C_{n}\varphi(p,0)$, in which $C_{n}$ indicates probability amplitude of the electronic state $|n\rangle$, $n=g$, $e$. We consider the Gaussian distribution function $\varphi(p,0)=\frac{1}{\pi^{1/4}\sqrt{\Pi}}e^{-(p-p_{c})^{2}/2\Pi^{2}}$ for the initial atom wave packet $\varphi(p,0)$. Here, $p_{c}$ is the center momentum and $\Pi$ represents the characteristic momentum, describing the position and width of wave packet in momentum space, respectively. In the following numerical treatment, $^{87}$Sr atom is characterized by its mass $M=86.9088775u$ ($u=1.67\times10^{-27}$ kg) and a long coherent time clock transition $^{1}$S$_{0}-^{3}$P$_{0}$ with frequency $429228004229873.65$ Hz~\cite{Campbell}.

\begin{figure}
  \includegraphics[width=8 cm]{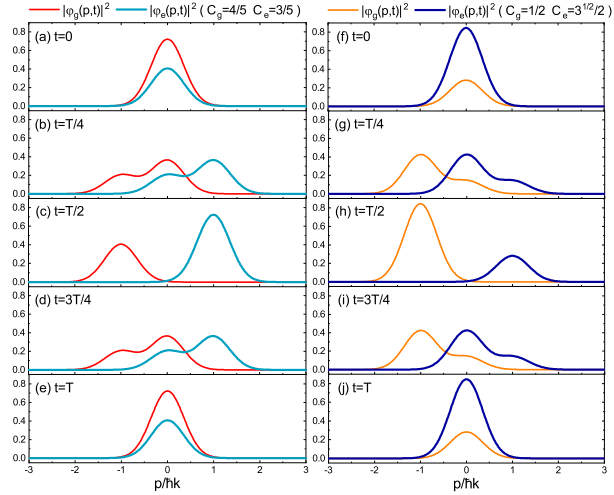}\\
  \caption{Time evolution of probability distributions in momentum space during a periodicity for two different initial states. The corresponding parameters are $\Pi=0.5\hbar k$, $\Omega=2\pi\times1$ MHz, $\Delta=0$ and $p_{c}=0$.}
  \label{fig3}
\end{figure}

\begin{center}
\textbf{3. Coherent evolution of the wave packet in momentum space}
\end{center}

For the convenience of following discussion, we write the wave function \eqref{eq:solution1} in the form of momentum spectrum. It can be achieved by replacing $p$ in $\int dp \varphi_{e}(p+\hbar k,t)|e,p+\hbar k\rangle$ of Eq.\eqref{eq:solution1} with $p-\hbar k$. Then, we have wave function in the momentum spectrum,
\begin{eqnarray}
|\varphi(t)\rangle=\int dp (\varphi_{g}(p,t)|g,p\rangle+\varphi_{e}(p,t)|e,p\rangle),
\label{eq:solution2}
\end{eqnarray}
For $t=0$, Eq.\eqref{eq:solution2} naturally gives rise to the initial state
\begin{eqnarray}
|\varphi(0)\rangle=(C_{g}|g\rangle+C_{e}|e\rangle)\int dp \varphi(p,0)|p\rangle,
\label{eq:initial-state}
\end{eqnarray}
in which the atom internal state is decoupled to momentum of the center of mass.

Based on Eq.\eqref{eq:solution2}, time evolutions of the atom wave packets in momentum space are illustrated in Fig.~\ref{fig3}. Here, a relatively stronger atom-light coupling has been considered, in which $\Omega$ is large enough that the relation $\Omega\gg\delta$ could be satisfied. It leads to the fact $\Sigma\approx\Omega$. In the strong coupling regime, the atom wave packet oscillates in momentum space with periodicity $T=\frac{2\pi}{\Omega}$. As shown in Fig.~\ref{fig3}, the atom wave packet can be seen to be composed of two parts, namely, $\varphi_{g}(p,t)$ and $\varphi_{e}(p,t)$. In this figure, we set different probability amplitudes to in the initial states to distinguish the two parts of the wave packet. In Fig.~\ref{fig3} (a)-(e), from time $t=0$ to $t=T/2$, wave packet in the part $\varphi_{g}(p,t)$ transits from the ground state to the excited state by gaining a negative momentum $-\hbar k$. On the other side, wave packet in the part $\varphi_{e}(p,t)$ transits from the excited state to the ground state by gaining a positive momentum $\hbar k$. During the consequent half period $T/2$, wave packet in the part $\varphi_{g}(p,t)$ transits back to the ground state by gaining a positive momentum $\hbar k$. At the same time, wave packet in the part $\varphi_{e}(p,t)$ transits back to the excited state by gaining a negative momentum $-\hbar k$. Similar process occurred in Fig.~\ref{fig3} (f)-(j), where a different initial state is considered.

\begin{center}
\textbf{4. Coherent evolution of the wave packet in position space}
\end{center}

\begin{figure}
  \includegraphics[width=8.5 cm]{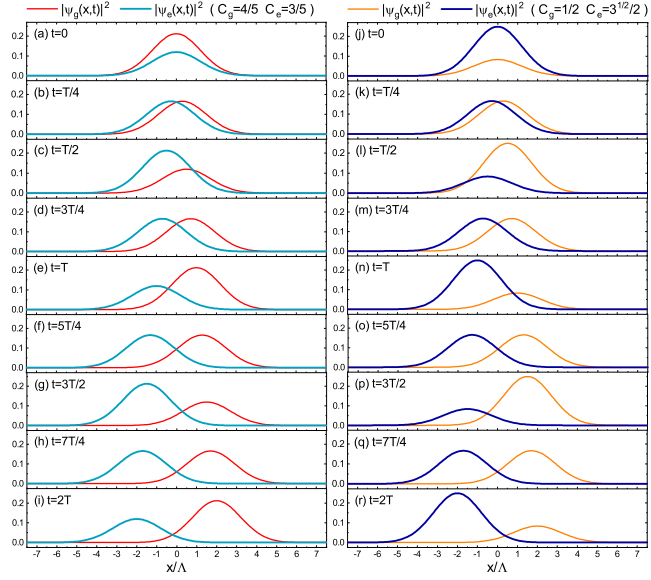}\\
  \caption{Time evolution of probability distributions in position space for two different initial states, where $\Pi=20\hbar k$, $\Omega=2\pi\times1$ MHz and $p_{c}=0$.}
  \label{fig4}
\end{figure}

Wave function in real position space $|\psi(x,t)\rangle$ can be derived through the Fourier transformation $|\psi(x,t)\rangle=\langle x|\varphi(t)\rangle$ in which $|\varphi(t)\rangle$ is given by Eqs.\eqref{eq:solution2}. Then, we have the wave function at position $x$ and time $t$,
\begin{eqnarray}
|\psi(x,t)\rangle=\psi_{g}(x,t)|g\rangle+\psi_{e}(x,t)|e\rangle.
\label{eq:wave-func}
\end{eqnarray}
The rigorous derivation is hard to give analytic results of these amplitudes $\psi_{g}(x,t)$ and $\psi_{e}(x,t)$. Fortunately, under the strong coupling regime $\Sigma\approx\Omega$, one could reach
\begin{eqnarray}
\psi_{g}(x,t)=\frac{\sqrt{\Pi}}{\pi^{1/4}\sqrt{\hbar}}[C_{g}\cos(\frac{\Omega t}{2})+iC_{e}\sin(\frac{\Omega t}{2})e^{-ik[x-x_{+}(t)]}]e^{-\frac{\Pi^{2}}{2\hbar^{2}}[x-x_{+}(t)]^{2}}e^{\frac{i}{\hbar}(p_{c}x-\zeta_{g}t)}
\label{eq:wave-func1}
\end{eqnarray}
and
\begin{eqnarray}
\psi_{e}(x,t)=\frac{\sqrt{\Pi}}{\pi^{1/4}\sqrt{\hbar}}[C_{e}\cos(\frac{\Omega t}{2})+iC_{g}\sin(\frac{\Omega t}{2})e^{ik[x-x_{-}(t)]}]e^{-\frac{\Pi^{2}}{2\hbar^{2}}[x-x_{-}(t)]^{2}}e^{\frac{i}{\hbar}(p_{c}x-\zeta_{e}t)}.
\label{eq:wave-func2}
\end{eqnarray}
In above equations, the state dependent displacement is $x_{\pm}(t)=(\frac{p_{c}}{M}\pm\upsilon)t$, where $\upsilon=\frac{\hbar k}{2M}$. In addition, we have $\zeta_{g}=\frac{\hbar\Delta}{2}+\frac{p_{c}^{2}}{2M}+p_{c}\upsilon+\frac{\hbar^{2}k^{2}}{4M}$ and $\zeta_{e}=\frac{\hbar\Delta}{2}+\frac{p_{c}^{2}}{2M}-p_{c}\upsilon+\frac{\hbar^{2}k^{2}}{4M}$. Here, $\frac{p_{c}^{2}}{2M}$ and $p_{c}\upsilon$ are kinetic energy and Doppler shift with respect to the center momentum $p_{c}$. Eqs.\eqref{eq:wave-func1} and \eqref{eq:wave-func2} are the main results of this work. In the derivation of the wave function amplitudes $\psi_{g}(x,t)=\int dp \varphi_{g}(p,t)\langle x|p\rangle$ and $\psi_{e}(x,t)=\int dp \varphi_{e}(p,t)\langle x|p\rangle$, there is a hardly integrable time dependent term $e^{-\frac{it}{\hbar}\frac{(p-p_{c})^{2}}{2M}}$ in the Fourier transformation. The fast oscillating part of this term, for $p$ far from the center momentum $p_{c}$, should not have contribution to the wave function amplitudes. Considering this effect, it is taken to be $e^{-\frac{it}{\hbar}\frac{(p-p_{c})^{2}}{2M}}=1$ here, which is satisfied for $p$ near the value $p_{c}$.

The schematic illustration of the wave packet splitting in Fig.~\ref{fig1} (b) can be understood from the wave function evolution shown in Fig.~\ref{fig4}. The figures reveal atom internal states determine probability distributions of wave packets which are propagating on the $x$ coordinate. Behaviors of the wave packet in real space fit with evolution of the wave function in momentum space plotted in Fig.~\ref{fig3}. $|\psi_{g}(x,t)|^{2}$ and $|\psi_{e}(x,t)|^{2}$ are probability distributions of two superposition states. Atom is fluctuating between the two states with the periodicity $T$. After one periodicity $T$, the wave packets move a distance $\Lambda=\frac{\pi\hbar k}{M\Omega}$ which is a tunable step length here as it is related to the parameters of light and atoms. If one detects position of the walking atom after an integer number $N$ of periodicity $T$, the atom should be at the position $N\Lambda$ but also at the position $-N\Lambda$. Probabilities of the detection would be determined by the internal electronic state of the atom.

\begin{figure}
  \includegraphics[width=6 cm]{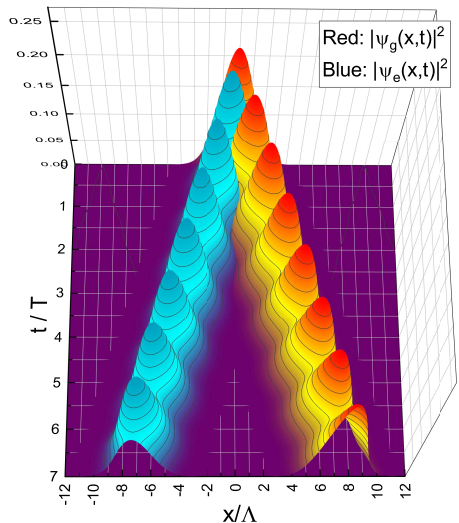}\\
  \caption{Deflection of the single atom wave packet with respect to the initial state $(\frac{4}{5}|g\rangle+\frac{3}{5}|e\rangle)\int dp\varphi(p,0)|p\rangle$, where $\Pi=20\hbar k$, $\Omega=2\pi\times1$ MHz, $\Delta=0$ and $p_{c}=0$.}
  \label{fig5}
\end{figure}

Deflection of the single atom wave packet can be seen more clearly in Fig.~\ref{fig5}. One can see wave peaks after every half periodic time $T/2$, which are corresponding to larger wave packets appeared in Fig.~\ref{fig4} (a). In contrast, the wave troughs in Fig.~\ref{fig5} are corresponding to the smaller wave packets in Fig.~\ref{fig4} (a). Although the phenomenon is analogous to the optical Stern-Gerlach effect observed in earlier experiment with laser standing wave~\cite{Sleator}, it is should be pointed out that the present process is occurred in the model with single traveling light but not standing waves. The coherent absorbtion and emission of photons are significant for the traveling light based Stern-Gerlach effect.

From perceptual intuition, the displacements of atom wave packets along $x$ and $-x$ coordinates in Fig.~\ref{fig4} and ~\ref{fig5} are similar to quantum walks~\cite{Aharonov}. Quantum walks can be realized with optical lattices~\cite{Dur}, single atoms~\cite{Karski}  and trapped ions~\cite{Zahringer}. However, actually there is difference that the probabilities of wave packet transitions in $x$ and $-x$ coordinates here are determined by initial state of the atom. It means the wave packet transitions during the splitting are deterministic. The quantum walks always indicates quantum random walks in which transitions of a quantum particle are randomly determined by a corresponding quantum coin at every step of transition~\cite{Aharonov,Dur,Karski,Zahringer}. In other words, transitions in quantum walks do not have to do with initial states.

In the intensity gradient induced optical Stern-Gerlach effect~\cite{Kazantsev,Sleator}, atom splitting can be observed when atoms resonantly interact with inhomogeneous optical field, in which dipole moment gives rise to fluctuating gradient force along the direction of the optical field gradient. In our model, optical field is allowed to be homogeneous. Fluctuating force acting on an atom comes from the coherent process of momentum gain and loss during the stimulated absorption and stimulated emission of photons accompanied by atom oscillation between two internal states. Furthermore, the fluctuating force is in the direction of light propagation. In the intensity gradient induced optical Stern-Gerlach effect~\cite{Kazantsev,Sleator}, an atomic beam in a strong resonant field splits into two beams of equal intensity. In the present traveling-wave light associated optical Stern-Gerlach effect, we will show that an input atomic beam splits into two beams with intensity that determined by the occupation distributions of atom internal states. In our model, even a single atom wave packet could splits into the superposition of two wave packets.

\begin{center}
\textbf{5. Spin-orbit coupling}
\end{center}

In Fig.~\ref{fig4}, the ground state part of the atom is in fact moving along the $x$ coordinate and the excited state part of the atom is moving along the opposite direction, i.e., the $-x$ direction. It is originated from the spin-orbit coupling in coherent atom-light interaction which has been observed previously in experiments using two or three Raman lasers~\cite{Lin,Curiel}. To illustrate the spin-orbit coupling effect quantitatively, let us further analyze Eq.\eqref{eq:S-equ2}. We translate the momentum coordinate by $\hbar k/2$ in Eq.\eqref{eq:S-equ2}. It is equivalent to replace the momentum $p$ in Eq.\eqref{eq:S-equ2} with $p_{x}-\frac{\hbar k}{2}$, then the equation would be changed into a symmetrical form,
\begin{eqnarray}
i\hbar\frac{\partial}{\partial t}\left[\begin{array}{c}
     \varphi_{g}(p_{x}-\frac{\hbar k}{2},t) \\
     \varphi_{e}(p_{x}+\frac{\hbar k}{2},t)
  \end{array}\right]=\left[\begin{array}{c c}
      \omega_{p_{x}-\frac{\hbar k}{2}} & -J \\
     -J & \omega_{p_{x}+\frac{\hbar k}{2}}
  \end{array}\right]\left[\begin{array}{c}
     \varphi_{g}(p_{x}-\frac{\hbar k}{2},t) \\
     \varphi_{e}(p_{x}+\frac{\hbar k}{2},t)
  \end{array}\right],
\label{eq:S-equ3}
\end{eqnarray}
where we have set the detuning to be $\Delta=0$ and $p_{x}$ is the momentum in $x$ coordinate. Considering expression of $\omega_{p_{x}\pm\frac{\hbar k}{2}}$ mentioned in Eq.\eqref{eq:S-equ2}, the Hamiltonian of Eq.\eqref{eq:S-equ3} in detail is
\begin{eqnarray}
V(p_{x})=\left[\begin{array}{c c}
      \frac{p_{x}^{2}}{2M}-\frac{\hbar k p_{x}}{2M}+\frac{\hbar^{2} k^{2}}{8M} & -J \\
     -J & \frac{p_{x}^{2}}{2M}+\frac{\hbar k p_{x}}{2M}+\frac{\hbar^{2} k^{2}}{8M}
  \end{array}\right].
\label{eq:V-Ham}
\end{eqnarray}
Now, we use standard Pauli matrixes to rewrite Hamiltonian \eqref{eq:V-Ham} as
\begin{eqnarray}
V(p)=(\frac{p_{x}^{2}}{2M}+\frac{1}{4}\hbar \omega_{B}) I-\upsilon p_{x}\sigma_{z}-J\sigma_{x},
\label{eq:V-Ham2}
\end{eqnarray}
where $\upsilon=\frac{\hbar k}{2M}$ is the velocity given in Eqs.\eqref{eq:wave-func1} and \eqref{eq:wave-func2}. The second term $-\upsilon p_{x}\sigma_{z}$ indicates spin-orbit coupling and the third term $-J\sigma_{x}$ is effective Zeeman splitting~\cite{Meng,Zheng}. In addition, $I$ represents $2\times2$ identity matrix, $\sigma_{x}$ and $\sigma_{z}$ are pauli matrixes. For small $p_{x}\rightarrow 0$, Eq.\eqref{eq:V-Ham2} would be reduced into a Hamiltonian which linearly depends on momentum $p_{x}$,
\begin{eqnarray}
V(p)=\frac{1}{4}\hbar \omega_{B} I-\upsilon p_{x}\sigma_{z}-J\sigma_{x}+O(p_{x}^{2}).
\label{eq:V-Ham3}
\end{eqnarray}
When the higher order term $O(p_{x}^{2})$ is ignored, Eq. \eqref{eq:V-Ham3} becomes Dirac like Hamiltonian~\cite{Meng,Huang}. One can observe Dirac cone near the point $p_{x}=0$ $(p=-\frac{\hbar k}{2})$ in Fig.~\ref{fig2} (a).

\begin{figure}
  \includegraphics[width=8 cm]{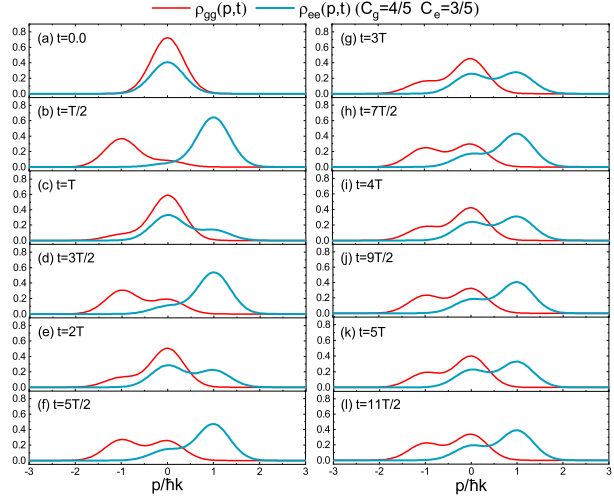}\\
  \caption{Evolution of the atom wave packet in momentum space with atom decay state $\Gamma=0.1\Omega$. $(\frac{4}{5}|g\rangle+\frac{3}{5}|e\rangle)\int dp\varphi(p,0)|p\rangle$, where $\Pi=0.5\hbar k$, $\Omega=2\pi\times1$ MHz, $\Delta=0$ and $p_{c}=0$.}
  \label{fig6}
\end{figure}

\begin{center}
\textbf{6. Influence from atom decay}
\end{center}
To consider the influence from atom relaxation and decoherence to the optical Stern-Gerlach effect, we write Eq.\eqref{eq:S-equ2} into the density matrix form as
\begin{eqnarray}
i\dot{\rho}_{gg}(p,t)=J\rho_{ge}(p,t)-J\rho_{eg}(p,t),
\label{eq:DensityMatrix1}
\end{eqnarray}
\begin{eqnarray}
i\dot{\rho}_{ee}(p+\hbar k,t)=-J\rho_{ge}(p,t)+J\rho_{eg}(p,t),
\label{eq:DensityMatrix2}
\end{eqnarray}
\begin{eqnarray}
i\dot{\rho}_{ge}(p,t)=-\delta\rho_{ge}(p,t)-J\rho_{ee}(p+\hbar k,t)+J\rho_{gg}(p,t),
\label{eq:DensityMatrix3}
\end{eqnarray}
where $\rho_{gg}(p,t)=|\varphi_{g}(p,t)|^{2}$, $\rho_{ee}(p+\hbar k,t)=|\varphi_{e}^{*}(p+\hbar k,t)|^{2}$, $\rho_{ge}(p,t)=\varphi_{g}(p,t)\varphi_{e}^{*}(p+\hbar k,t)$ and $\rho_{eg}(p,t)=\rho_{ge}^{*}(p,t)$. Relaxation and decoherence of atoms are generally originated from coupling between the atoms and their environments. Here, we consider a zero temperature thermal reservoir as the environment~\cite{Scully}. At the same time, it is assumed that the reservoir is only coupled to internals states $|g\rangle$ and $|e\rangle$ of the atoms. According to the quantum theory of damping~\cite{Scully}, we phenomenologically induce the effect of atom decay. Then, Eqs. \eqref{eq:DensityMatrix1}-\eqref{eq:DensityMatrix3} could be rewritten in the form,
\begin{eqnarray}
i\dot{\rho}_{gg}(p,t)=i\Gamma \rho_{ee}(p+\hbar k,t)+J\rho_{ge}(p,t)-J\rho_{eg}(p,t),
\label{eq:DensityMatrix4}
\end{eqnarray}
\begin{eqnarray}
i\dot{\rho}_{ee}(p+\hbar k,t)=-i\Gamma \rho_{ee}(p+\hbar k,t)-J\rho_{ge}(p,t)+J\rho_{eg}(p,t),
\label{eq:DensityMatrix5}
\end{eqnarray}
\begin{eqnarray}
i\dot{\rho}_{ge}(p,t)=-(\delta+\frac{i\Gamma}{2})\rho_{ge}(p,t)-J\rho_{ee}(p+\hbar k,t)+J\rho_{gg}(p,t),
\label{eq:DensityMatrix6}
\end{eqnarray}
where $\Gamma$ represents atom decay rate which is identical to the damping rate achieved in the Weisskopf-Wigner theory of spontaneous emission~\cite{Scully}. Eqs. \eqref{eq:DensityMatrix4}-\eqref{eq:DensityMatrix6} obviously satisfy sum of the probabilities $\rho_{gg}(p,t)+\rho_{ee}(p+\hbar k,t)$ is a constant at any time.

\begin{figure}
  \includegraphics[width=8 cm]{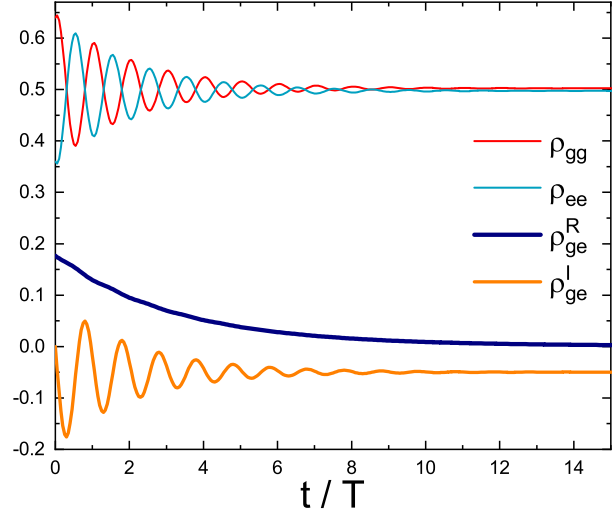}\\
  \caption{Atom distribution probabilities in the two levels and the coherent term as a function of time. $(\frac{4}{5}|g\rangle+\frac{3}{5}|e\rangle)\int dp\varphi(p,0)|p\rangle$, where $\Pi=0.5\hbar k$, $\Omega=2\pi\times1$ MHz, $\Gamma=0.1\Omega$, $\Delta=0$ and $p_{c}=0$.}
  \label{fig7}
\end{figure}

Initial state of the atom wave packet set in Fig.~\ref{fig6} is exactly the same as that in Fig.~\ref{fig5}(a)-(e). Furthermore, $\rho_{gg}(p,t)$ and $\rho_{ee}(p,t)$ are identical to $|\varphi_{g}(p,t)|^{2}$ and $|\varphi_{e}(p,t)|^{2}$. The difference is that we have considered atom decay with the rate $\Gamma$ in Fig.~\ref{fig6}. In the fully coherent system as shown in Fig.~\ref{fig5}, atom wave packet distributions after a period time $t=T$ can return to the form of initial time $t=0$. The system keeps its fluctuation between the states $\varphi_{g}(p,t)$ and $\varphi_{e}(p,t)$. However, time evolution of the momentum distribution in Fig.~\ref{fig6} could not return to the initial state forever. The atom wave packet would evolve to a stationary state. Fluctuation between the two states $|g\rangle$ and $|e\rangle$ tends to be disappeared when the time is longer than $4T$. Now, we calculate the probabilities of the atom at two internal states $\rho_{gg}(t)=\int |\varphi_{g}(p,t)|^{2}dp$, $\rho_{ee}(t)=\int |\varphi_{g}(p+\hbar k,t)|^{2}dp$ and the coherent term $\rho_{ge}(t)=\int \varphi_{g}(p,t)\varphi_{e}^{*}(p+\hbar k,t)dp$. As illustrated in Fig.~\ref{fig7}, due to the thermal reservoir induced decoherence, the atom is distributed on the states $|g\rangle$ and $|e\rangle$ with almost half probability, respectively. Evidence of the decoherence is that real part $\rho_{ge}^{R}$ of the coherent term is close to be zero after a long enough time. However, the imaginary part $\rho_{ge}^{I}$ of the coherent term is a nonzero quantity. They can be further proved by stationary solutions of Eqs. \eqref{eq:DensityMatrix4}-\eqref{eq:DensityMatrix6}. When one integrate these equations with respect to momentum $p$, they become equations of the probabilities $\rho_{gg}(t)$, $\rho_{ee}(t)$ and $\rho_{ge}(t)$. Considering the normalization condition $\rho_{gg}+\rho_{ee}=1$, one obtains the stationary solutions as follows
\begin{eqnarray}
\rho_{ee}=\frac{\Omega}{\frac{4\delta^{2}}{\Omega}+\frac{\Gamma^{2}}{\Omega}+2\Omega},
\label{eq:solution3}
\end{eqnarray}\begin{eqnarray}
\rho_{ge}^{R}=\frac{2\delta}{\frac{4\delta^{2}}{\Omega}+\frac{\Gamma^{2}}{\Omega}+2\Omega},
\label{eq:solution1}
\end{eqnarray}
and
\begin{eqnarray}
\rho_{ge}^{I}=-\frac{\Gamma}{\frac{4\delta^{2}}{\Omega}+\frac{\Gamma^{2}}{\Omega}+2\Omega}.
\label{eq:solution2}
\end{eqnarray}
Since in our parameters, $\delta$, $\Gamma$ $\ll$ $\Omega$, these results are consist with the numerical outputs plotted in Fig.~\ref{fig7}.

\begin{center}
\textbf{7. Influence from atom-atom interactions}
\end{center}

Until now, we have considered the dynamics of single-atom supposing the atom gas is diluted. In practical Stern-Gerlach type experiment, many-atom process would be involved and atom-atom interactions may affect the atomic beam splitting effect. Except inter atom interaction, an atom even can interact with itself as a quantum particle.

For cold atoms as we considered here, inter atom interactions are characterized by repulsive forces~\cite{Dalfovo,Pawlowski,Escriva}. Atom-atom action such as Casimir-Polder interaction decays exponentially and tends to disappear within inter atom distance $R\sim1$ $\mu$m~\cite{Bostrom}. The Van der Waals interaction decays with the order $1/R^{6}$ and tends to disappear within the inter atom distance $R\sim2$ $\mu$m~\cite{Johnson}. On contrast, dipole-dipole interaction varies as $1/R^{3}$ and be going to disappear around the distance $R\sim3$ $\mu$m~\cite{Johnson}. When density of atomic gas is large enough, atom-atom interactions should interrupt the periodic walking of atom wave packet shown in Fig.~\ref{fig4} and ~\ref{fig5}. The inter atom interactions may also cause dephasing to the internal states of atoms due to the random collisions~\cite{Horikoshi}. Obviously, when density of the atomic gas is smaller than the estimated value $1/R^{3}$, these interaction could be neglected. The threshold density for the Casimir-Polder interaction is $10^{12}/$ cm$^{3}$, for the Van der Waals interaction is $2.5\times10^{11}/$ cm$^{3}$ and for the dipole-dipole interaction is $1.1\times10^{11}/$ cm$^{3}$. Considering multi levels of atoms involved in the optical couplings, entanglement between two neighboring atoms could be realized with optical pulses~\cite{Lukin,Leseleuc}. Therefore, based on the atom wave packet superposition states described in our present model, to realize entanglement of multi-atom wave packets could be predicted here. Since motion of the wave packets can be controlled through the atom internal states associate with optical fields.

For a single atom, when its wave packet is divided into two components, there would be interaction between the two components due to atom-atom interaction effect. Therefore, a single atom can interact with itself. Under above parameters, value of the wave packet velocity is $\upsilon=2.92\times10^{3}$ $\mu$m/s. It means the single atom wave packet splitting is influenced by the atom-atom interactions before about $0.001$ s. During this time, the inter atom repulsive forces should promote the wave packet splitting. After distance between the two components of the wave function becomes longer than $3$ $\mu$m, the influence of atom-atom interaction between the two components would gradually disappear.

\begin{figure}
  \includegraphics[width=7 cm]{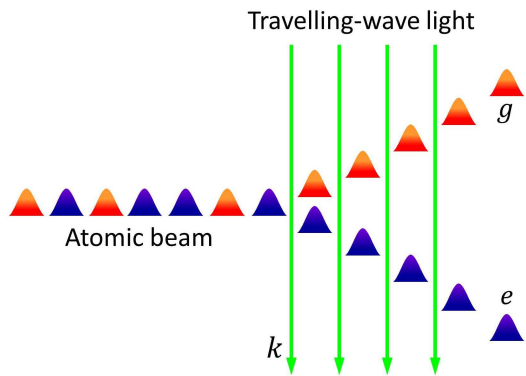}\\
  \caption{An input atomic beam is perpendicular to a traveling-wave light and splits into two beams. $k$ is the wave vector of the light and $g$ ($e$) represents wave packet corresponding to the ground state (the excited state).}
  \label{fig8}
\end{figure}

\begin{center}
\textbf{8. Discussion about experimental justifications of this model}
\end{center}
Now, let discuss experimental feasibility of the traveling-wave light induced optical Stern-Gerlach effect. Coherent coupling between two-level atoms and a laser field could be satisfied using the optical clock transitions of alkali-earth (like) atoms, which are desired for our model. Especially, coherent properties of the $^{1}$S$_{0}-^{3}$P$_{0}$ transition of Sr atoms have been measured recently with nondestructive measurements~\cite{Muniz}. It is reported that lifetime of the transition reaches from hundreds milliseconds to a few hours. Long lifetimes of the excited state $^{3}$P$_{0}$ have been reported in other experiments with different methods~\cite{Dorscher,Porsev}.

In optical lattice, spin-orbit couplings are created using optical clock transitions of alkali-earth (like) atoms~\cite{Livi}. The spin-orbit coupling effect of cold atoms observed in the optical lattices is equivalent to that in our present process. The main difference is that our atoms are in free-space but not trapped in lattice potentials. Coupling between atomic internal states and its momentum of center of mass play important role in the chiral current of atomic transport in optical lattices~\cite{Michael}. Actually, this is also the key mechanism of our system configuration.

Atomic beam could be created with a directional oven and cooled down consequently to ultra-low temperature as described in a recent Stern-Gerlach experiment~\cite{Melin}. The best direction of the incident atomic beam in our system should be perpendicular to the propagation of traveling-wave light as conceptually shown in Fig.~\ref{fig8}.

We investigate our model based on $^{87}$Sr atom in this work. Its long coherent clock transition is significant for the model. Actually, other neutral atoms, especially the alkaline-earth (like) atoms which are characterized by long coherent optical transitions, such as Li, Na, K, Rb, Cs, Mg and Ca, should be applied to observe the optical Stern-Gerlach effect. As derived in Eqs.\eqref{eq:wave-func1} and \eqref{eq:wave-func2}, the velocity $\upsilon=\frac{\hbar k}{2M}$ of atom wave packet depends on atom transition wave length and the atom mass. Therefore, in principle, different atoms in a beam have different velocities along the direction of light propagation.

\begin{center}
\textbf{9. Conclusions}
\end{center}

In conclusions, we proposed a simplified model of optical Stern-Gerlach effect under coherent coupling between single atoms and a traveling light. The chiral motion of the atom deflection in this effect can be connected to spin-orbit coupling appeared in the coherent interaction. The coherence is generated with strong atom-light coupling, removing the noises from Doppler effect and aback action. In this process, the probability amplitudes of atom wave packets superposition state depend on the initial internal state of the atom. In addition, step length of the atom walking can be tuned arbitrarily by changing the corresponding Rabi frequency for given atoms. Analytical results in this work are suitable for the case where momentum uncertainty of an individual atom is very narrow along the direction of light propagation. The traveling light based optical Stern-Gerlach effect may have important applications for the realization of nanometer to micrometer scaled atom beam splitter, atom interferometry, and neutral atom based quantum computation.

\begin{acknowledgments}
This work was supported by R \& D  Program of Beijing Municipal Education Commission (KM202011232017).
\end{acknowledgments}

\end{document}